\documentclass[aps,prl,twocolumn,groupedaddress,amssymb]{revtex4}
\usepackage{graphicx}
\usepackage[figuresright]{rotating}
\begin{document}

{\bf Comment on "Why Do Gallium Clusters Have a Higher Melting
Point than the Bulk?"}

The computational work~\cite{joshi3}, motivated by recent
experiments~\cite{breaux} on the heat capacity measurements of
small Ga$_{39}^+$ and Ga$_{40}^+$ clusters, claimed that the
observed broad peak in the heat capacity represents melting of the
small size clusters and made a strong point that due to the
special character of the chemical bonds, these clusters, contrary
to all expectations, melt at temperatures higher than the
corresponding bulk material.  In order to understand mysterious
``higher-than-bulk melting temperatures" in small gallium
clusters, Ga$_{17}$ and Ga$_{13}$, they utilized the powerful
machinery of the density functional theory (DFT) molecular dynamic
(MD) simulations in a form of the isokinetic Born-Oppenheimer MD,
using ultrasoft pseudopotentials within the LDA. The specific-heat
curve, calculated by the multiple-histogram technique, indeed
showed the peak in the heat capacity to be well above the bulk
melting point of 303 K, viz., around 650 and 1400 K for Ga$_{17}$
and Ga$_{13}$, respectively. The ``higher-than-bulk melting
temperatures" were attributed mainly to the covalent bonding in
clusters, contrasting the covalent-metallic bonding in the bulk.

In our Comment we show that the peak in the heat capacity is in
fact a well known generic behavior of finite size systems usually
referred to as {\it Schottky anomaly}.  Thus the connection of the
peaks calculated
 in~\cite{joshi3} (and also in the earlier works
~\cite{joshi1,joshi2}) to the real world is questionable.

The thermodynamic behavior of a finite system consisting of $N$
particles is well discussed (as an exercise) in the R. Kubo
textbook~\cite{kubo}, Chapter 1, Example 4 on pages 38-41. When a
system contains a substance having the excitation energy $\Delta
E$, the specific heat is given by the formula (9) of Example 4:
$$
C=Nk_B \left(\frac{\Delta E}{k_BT}\right)^2
{\exp\left(\frac{\Delta E}{k_BT}\right)}\left/
{\left(1+\exp\left(\frac{\Delta E}{k_BT}\right)\right)^2}\right. .
$$
\begin{figure}
\centering
\includegraphics[angle=0,width=1.8in,height=1.9in]{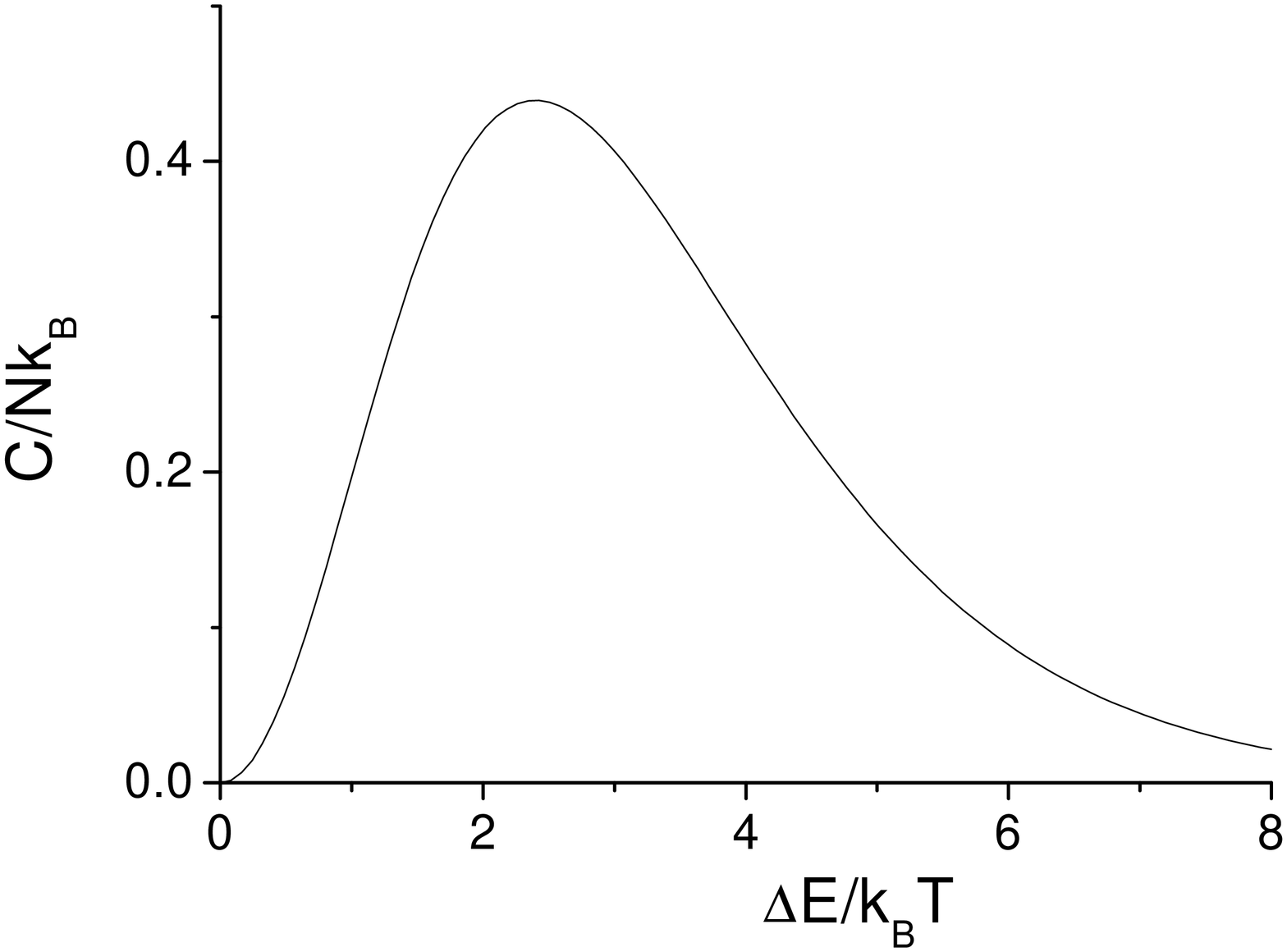}~\includegraphics[angle=0,width=1.8in,height=1.9in]{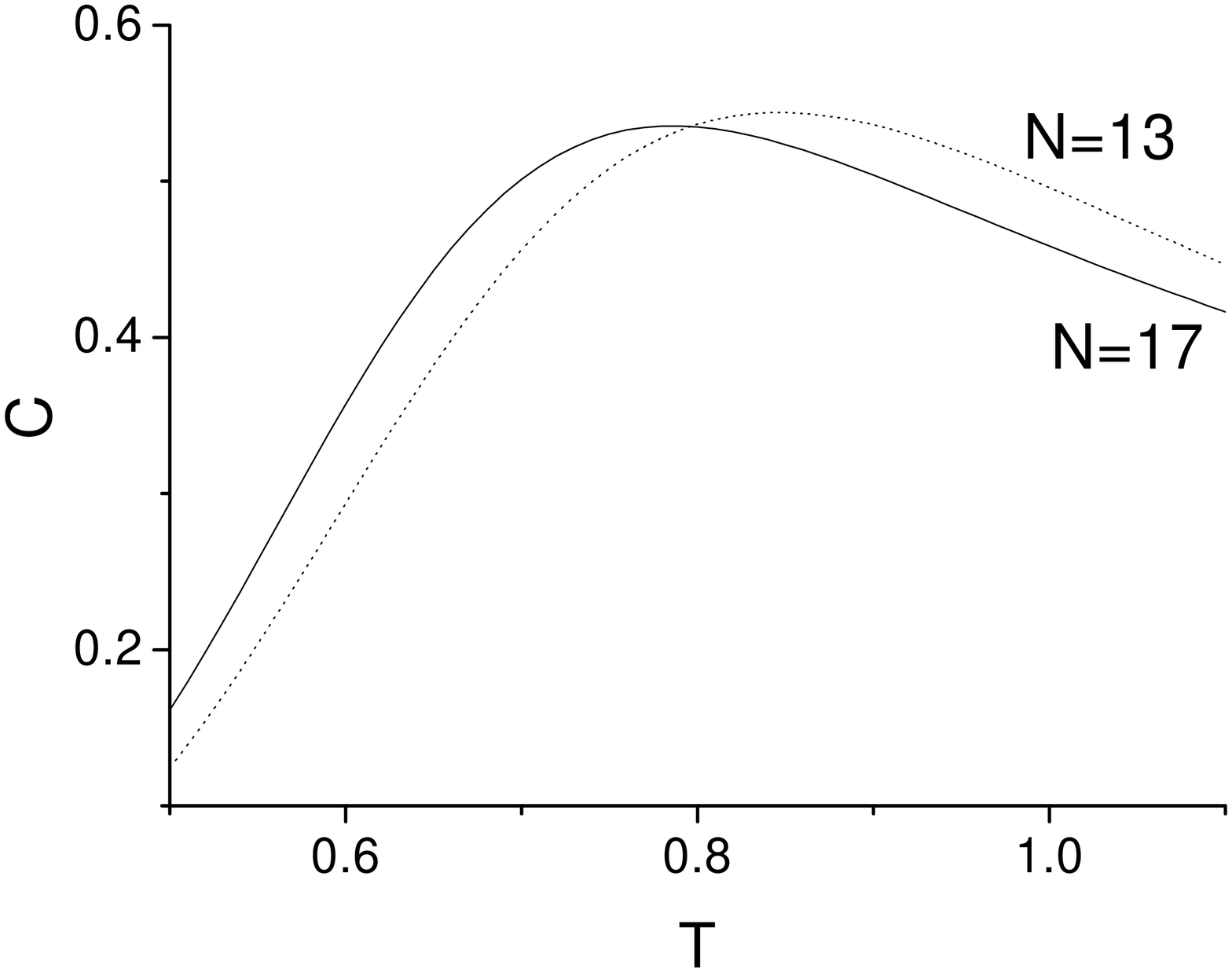}
\caption{Left: The Schottky specific heat. Figure~1.9 from~\cite{kubo}.
Right: Specific heat $C_V$ per spin of the Ising spin chain with
$N=17$ (solid line) and $N=13$ (dotted line) spins.}
\end{figure}
\noindent Shown in Figure~1 is the corresponding peak.  The textbook
exercise refers to the system of noninteracting particles.  To see
qualitatively what do interactions contribute to Schottky anomaly,
one can consider, for example, a finite system of interacting spins.
To avoid unnecessary complications we take the simplest analytically
solvable model for a spin chain closed into the ring, i.e. the
one-dimensional Ising model with the periodic boundary
conditions~\cite{baxter}. The specific heat per spin
$$
C=\frac{\partial}{\partial T}\left(\frac{k_BT^2}{N}\frac{\partial}{\partial
 T} \log(\lambda_+^N+\lambda_-^N)\right), \;\;
\lambda_\pm=e^\frac{J}{k_BT}\pm e^\frac{-J}{k_BT},
$$
\noindent ($J$ is the coupling constant) is shown in Figure~1. Note
the shift of the position of the maximum to higher temperatures
and growth of the maximum magnitude for the smaller cluster size
exactly as presented in Figs.~4 and~5 of~\cite{joshi3}.  This kind
of the size dependence is indeed typical for the Schottky anomaly.
Interactions thus can only slightly change its shape, but the peak
itself remains intact.  This concludes our demonstration that the
broad maximum in the heat capacity observed in \cite{breaux}
should be attributed rather to Schottky anomaly, than to cluster
melting.

 This research is supported by the US DOE Office of
Science under contract No. W-31-109-ENG-38.

\noindent
Lev N. Shchur and Valerii M. Vinokur \\
Materials Science Division, Argonne National Laboratory, Argonne,
Illinois 60439, USA

\pacs{61.46.+w, 36.40.Cg, 36.40.Ei}

\end{document}